# Record High Ramping Rates in HTS Based Superconducting Accelerator Magnet

H. Piekarz, S. Hays, B. Claypool, M. Kufer and V. Shiltsev

*Abstract* We report results of experimental test of the High Temperature Superconductor (HTS) based fast-cycling prototype accelerator magnet capable to operate up to about 300 T/s field ramping rate with some 0.5 T B-field in the magnet gap. The measured upper limit for the cryogenic cooling power required to support magnet conductor operation at high ramping rates indicates great potential for such type of magnets in rapid cycling synchrotrons for neutrino research or muon acceleration. The test magnet design, construction and supporting cryogenic and power systems are briefly described. The magnet's power test results are discussed in terms of a possible upgrade of this magnet design to 2 T B-field, a maximum feasible with super-ferric magnet.

*Index Terms*— Superconducting accelerator magnets, Rapid cycling magnets, High Temperature Superconductors (HTS).

## I. HTS RAPID-CYCLING ACCELERATOR MAGNET DESIGN

THE AC losses in the fast-cycling magnet are due to power losses in both magnet energizing conductor and magnetic core. The power losses in the magnetic core can be reduced by using as thin as practically possible the Fe(n%Si) laminations. The power losses in the conductor can be reduced by minimizing both its mass and the exposure area to ramping magnetic field descending from the core. Application of the superconducting power cable allows to strongly minimize its mass and size, and as a result also the size and mass of the magnetic core. In addition, use of the HTS conductor allows significantly increase the operational temperature margin of power cable facilitating in this way temperature-based quench detection and protection system.

The magnet design and its first power test results at ramping rate of 12 T/s were presented earlier [1]. To facilitate discussion of the power test results reported here we outline briefly the previously reported technical information on this magnet design and construction. The conceptual design of the HTS based rapid-cycling test magnet is shown in Fig. 1. The 0.5 m long magnet core of 620 mm x 255 mm cross-section is of the shape as shown in Fig. 1 with two beam gaps of 100 mm (hor.). x 10 mm (vert.). A single magnet power cable energizes both beam gaps facilitating simultaneous acceleration of two beams. For the unipolar current wave-form B-fields in the beam gaps are of the opposite sign making it possible for the oppositely charged particle beams (such as $\mu^+$, $\mu^-$) to circulate in the same direction. Such an arrangement facilitates use of a common accelerating RF system for both beams. The 3-part core facilitates installation of the multi-turn magnet power coil. The superconducting power cable is assembled as a narrow structure placed within the magnet core space where magnetic field descending from the core is very minimal [1].

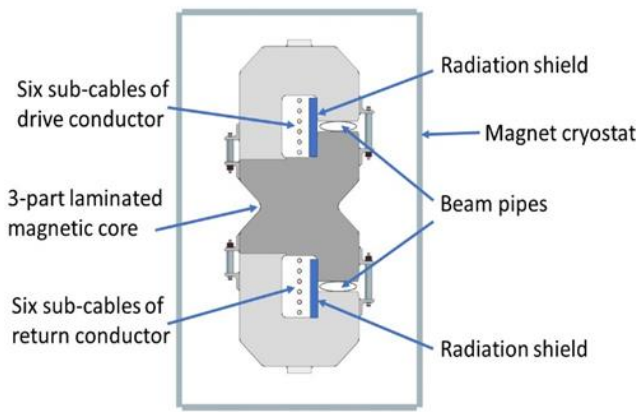

1. Accelerator magnet featuring 2 vertically arranged beam gaps energized by a single very narrow power cable assembled of multiple sub-cables

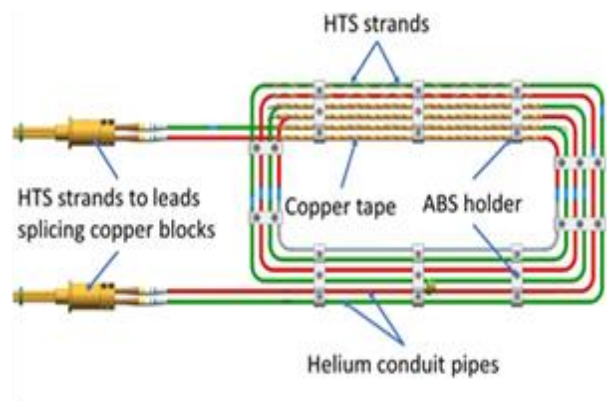

Fig. 2. Arrangement of 3-turn HTS conductor coil.

This work was supported by Fermi Research Alliance, LLC under contract No. De-AC02-07CH11359 with the United States Department of Energy. *(Corresponding author: Henryk Piekarz)*

H. Piekarz (e-mail: hpiekarz@fnal.gov), S. Hays, B. Claypool, M. Kufer and V. Shiltsev are with Fermi National Accelerator Laboratory, Batavia, IL 60510, USA.

Color versions of one or more of the figures in this paper are available online at http://ieeexplore.ieee.org. Digital Object Identifier will be inserted here upon acceptance.

Digital Object Identifier will be inserted here upon acceptance.

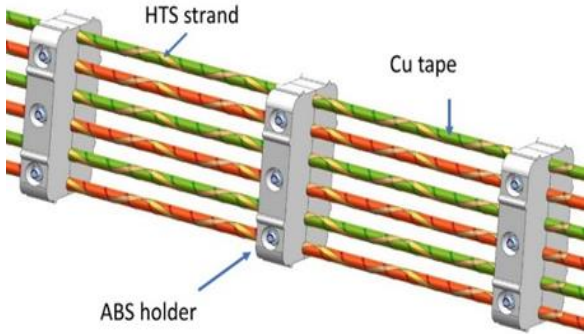

Fig. 3. Details of HTS conductor coil assembly

Two sub-cables are used to construct the 3-turn power coil (see Fig. 2). The more detailed view of conductor coil assembly is shown in Fig. 3. Two 2.5 mm wide and 0.1 mm thick HTS strands (Super-Power, Inc [3]) are helically wound at the 10 cm pitch on the surface of the 316LN helium conduit pipe of 8 mm ID, 0.5 mm wall. A single layer of 0.1 mm thick and 12.5 mm wide OFHC copper tape is wound helically over the strands to firmly secure their attachment to the helium conduit pipe. The ABS holders keep the conductor assembly together and isolating it from the magnet core. The 40-layer of MLI is wrapped over the cable structure between the ABS holders. Up to 10 HTS strands can be placed over each helium conduit pipe but for the test magnet there are only 2 strands per conduit pipe of 24 m length, of which 12 m is within the magnet core space. The mechanical properties of the HTS magnet ca-are listed in Table II. The cable critical current is 6 kA at 6 K and 2 kA at 30 K. Magnet was operated at maximum 1 kA current with the conductor at 6 K, preventing in this way a possibility of quench occurrence up to 30 K temperature rise.

TABLE I
HTS CABLE COMPONENTS WITHIN MAGNET CORE SPACE

| Cable component | Volume ($m^3$) | Mass (g) | Resistivity @ 5K ($\Omega$-m) |
|---|---|---|---|
| **HTS strands** | | | |
| Number of strands | 24 | | |
| REBCO | $2.4 \cdot 10^{-8}$ | 0.15 | |
| Hastello9y C-276 | $1.2 \cdot 10^{-7}$ | 10.7 | $1.2 \cdot 10^{-6}$ |
| Cu cup | $9.5 \cdot 10^{-7}$ | 8.6 | $8 \cdot 10^{-10}$ |
| Ag cup | $9.5 \cdot 10^{-8}$ | 1.0 | $1 \cdot 10^{-10}$ |
| Total strands | $22.5 \cdot 10^{-7}$ | 20.5 | |
| **Strand support** | | | |
| Cryogenic pipes | $73 \cdot 10^{-6}$ | 569 | $5 \cdot 10^{-7}$ |
| Copper tape | $15 \cdot 10^{-6}$ | 107 | $8 \cdot 10^{-10}$ |
| Total strands support | $88 \cdot 10^{-6}$ | 676 | |

TABLE II
ELECTRICAL PROPERTIES OF HTS MAGNET

| | |
|---|---|
| Cable critical current @ 6.5 K | 6 kA |
| Cable critical current @ 30 K | 2 kA |
| Magnet critical current @ 30 K | 6 kA |
| Magnet maximum B-field | 0.4 T |
| Magnet resistance @ 6.5 K with leads @ RT | 340 µΩ |
| Magnet resistance @ 6.5 K with cold leads | 150 µΩ |
| Magnet inductance | ~ 96 µH |
| Leads inductance | ~ 4 µH |

## II. MAGNET TEST ARRANGEMENT AND POWER TESTS

The magnet test system arrangement is shown in Fig. 4 and the actual test setup in Fig. 5. The magnet HTS conductor coil and the conventional (copper) current leads are cooled using the separate liquid helium flows. The temperature and pressure sensors are placed at the inlets and outlets of the liquid helium cooling circuits of the magnet conductor coil and the current leads. The measured difference of the conductor cooling helium temperature together with helium pressure is used to determine change of the helium enthalpy which in combination with measured helium flow rate gives the generated heat.

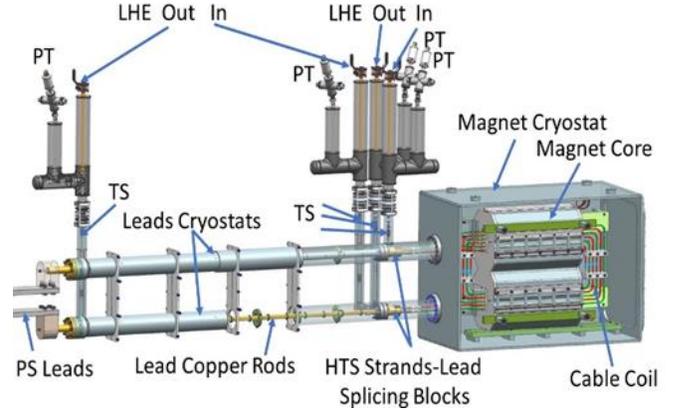

Fig. 4. HTS magnet test system arrangement: TS - temperature sensors, PT– pressure transducers

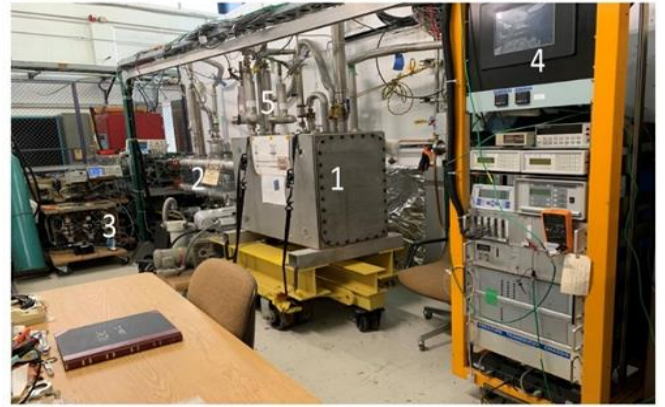

Fig. 5. View of the HTS magnet test setup: 1-magnet, 2-current leads, 3- power supply, 4- cryogenic controls, 5- LHe cryogenic lines.

The AC current source is based on the discharge of the capacitor bank constructed of twenty 960 µF capacitors charged up to 80 V. The discharge current of 1000 A, which for the 3- turn magnet conductor makes 3000 A to energize the magnet, generating about 0.4 T B-field in both beam gaps. The power supply operated in the unipolar and bipolar modes. In the unipolar mode the maximum B-field is 0.377 T, and in the bipolar mode the maximum B-field spans to 0.5 T. The full discharge current pulse length was 0.010 s with the peak magnetic field at about 0.003 s. The applied current pulse repetition rate was 1-10 Hz. The magnetic field and the dB/dt response for the unipolar and bipolar excitation currents is shown in Fig. 6 and Fig. 7, respectively. The maximum dB/dt rates with unipolar and bipolar current waveforms are 274 T/s and 289 T/s, respectively. These rates are obtained at the fraction of the maximum B-field. The dB/dt rates for the maximum B-fields are 160 T/s for the unipolar and 170 T/s for the bipolar current waveforms. The temperature change (dT) of the cooling liquid helium between the inlet and outlet of the magnet conductor coil with 1000 A excitation current and operating at 1 and 14 Hz repetition rates are shown in Fig and Fig. 9, respectively. With estimated temperature measurement error of +/- 0.003 K we conclude that the maximum temperature rise due to the increased repetition rate from 1 to 14 Hz, is less than 0.006 K. For the cooling liquid helium of 6 K, 0.28 MPa pressure and flow rate of 2.4 g/s this temperature change corresponds to the heat loss of less than 0.1 W. If the magnet conductor was not superconducting the heat dissipated in the copper tape holding strands would be 180 W, raising the liquid helium temperature to 39 K. The tentative analysis of the hysteresis and eddy losses in the HTS cable components, as listed in Table I, indicates that to be within the 0.1 W of the heat loss the cable would have to be exposed to the average B-field of about 0.02 T, or 5% of the 0.4 T in the beam gap. The cable component contributions to the heat loss are as follows: cryogenic pipes -70%, HTS strands – 19 % and Cu tape - 11%. The dominant power loss is due to the cryogenic pipes. The 0.5 mm wall of the 9 mm OD SS pipe used in the test magnet can be safely reduced by a factor of 5 to 0.1 mm wall as the allowable working pressure with thinner wall pipe will be 30 bar, strongly exceeding the conductor coil operational helium pressure of 3 bar. This allowable pressure creates sufficient safety margin for the operation of the quench detection and cable protection systems in the case of helium pressure rise due to the incoming magnet quench.

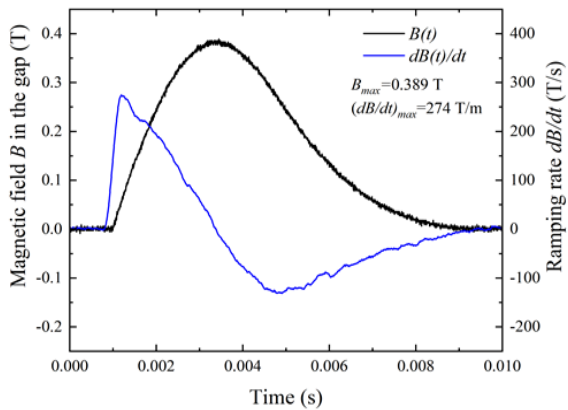

Fig. 6. Magnet B-field waveform with the unipolar current pulse

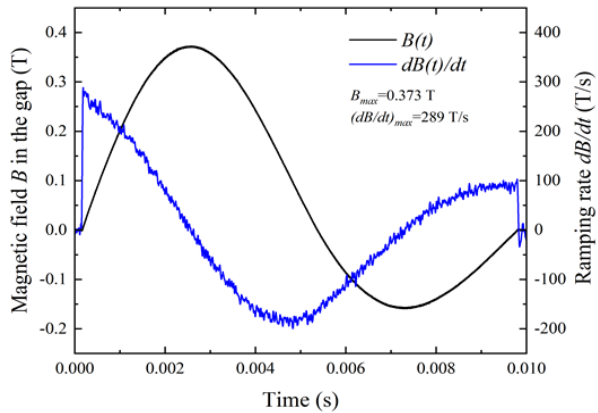

Fig. 7. Magnet B-field waveform with the bipolar current pulse

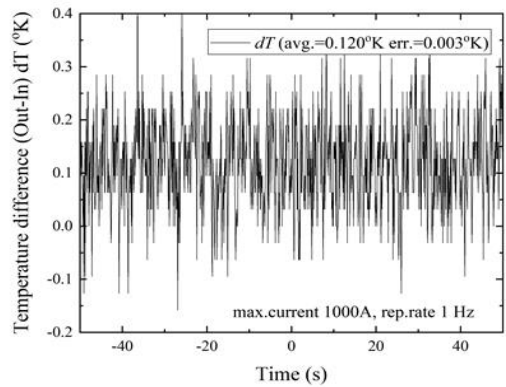

Fig. 8. Conductor coil temperature change at 1000 A and 1 Hz operation

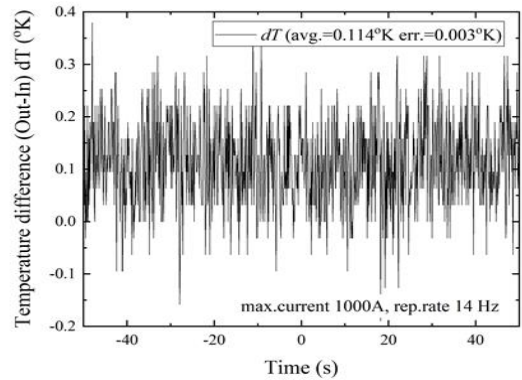

Fig. 9. Conductor coil temperature change at 1000 A and 14 Hz operation

## III. POSSIBLE APPLICATION OF HTS-BASED RAPID CYCLING MAGNET FOR THE MUON ACCELERATION

The achieved dB/dt rates with test HTS-based rapid cycling accelerator magnet open the possibility of using HTS magnet design in a future muon collider. In Ref. [2] a concept of the 14 TeV (c.m.e.) muon collider in the LHC tunnel is based on 16 T DC and 3.8 T fast-cycling (AC) magnet strings. The development of 16 T $Nb_3Sn$ accelerator magnets is underway by several groups worldwide. As more modest case below we consider using the existing LHC's 8 T DC magnets in combination with the 2 T rapid cycling HTS magnets based on the technology presented in this paper. Such an arrangement will allow for the construction of the muon collider scaled down to 7 TeV c.m.e. collision energy. It is important to point out, however, that the physics reach the $μ^+$- $μ^-$ collisions at 7 TeV c.m.e is equivalent to the 50 TeV c.m.e of the proton-proton collisions.

Particle acceleration requires linear response of the B-field in the magnet beam gap to the energizing current to have equal acceleration rumps on way to top energy. The magnetic core saturation in the super-ferric magnet sets the limit on the usable B-field range. The silicon steel laminations used for the HTS test magnet core fully saturate at about 2 T making the current to B-field linear response only up to 1.7 T. As an example, we show in Fig.10 a simulated B-field to current response for the 25 mm beam gap. Such a beam gap is considered for the muon accelerator [4] though the unipolar energizing current will be used with the magnet presented in this paper.

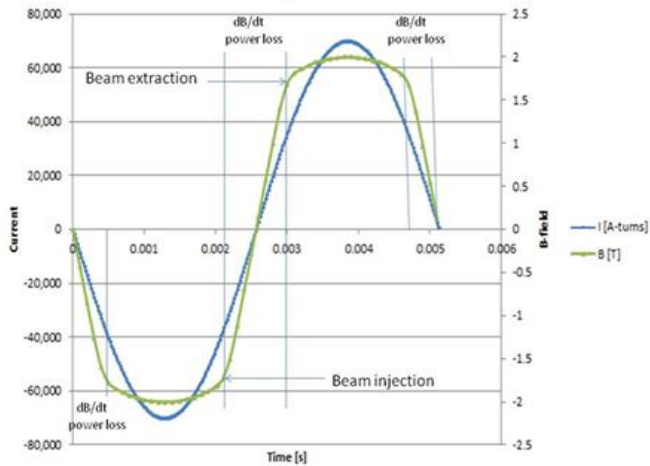

Fig. 10. B-field to current response with the full saturation at 2 T

Using the formulas from [2] the basic parameters of 7 TeV muon collider in LHC tunnel are listed in Table III. The dB/dt rate of 272 T/s for the HTS rapid cycling magnet has been achieved. We need to consider, however, feasibility of upgrading the test magnet design to 2 T B-field with 25 mm beam gaps. The test magnet conductor operational current is 6 kA at 30 K. By increasing the cold pipe diameter to 10 mm, and by adding a second layer of the HTS strands the magnet operational current can be increased 12-fold, or up to 72 kA-turn.

The HTS strands in the second layer will be wound at the opposite angle with respect to the strands in the first layer to minimize power loss caused by the self-field coupling between the layers of HTS strands. The 12-fold increase in the number of strands, however, will likely make strand's hysteresis and eddy losses to become dominant for the overall cable power loss.

TABLE III
PARAMETERS OF 7 TEV MUON COLLIDER IN LHC TUNNEL

| Accelerator parameter | |
|---|---|
| Circumference | 26.7 km |
| Maximum beam energy | 3.5 TeV |
| Injection energy | 0.45 TeV |
| DC magnet B-field | 8 T |
| AC magnet maximum B-field | 2 T |
| AC magnet linear response B-field | 1.7 T |
| AC magnet beam gap | 25 mm |
| AC magnet maximum current | 66 kA-turns |
| DC magnet string | 5.25 km |
| AC magnet string | 18.8 km |
| Accelerator ring filling factor | 0.9 |
| Number of turns to full energy | 351 |
| Energy gain per turn | 8.7 GeV |
| Repetition rate | 10 Hz |
| AC magnet current cycle | 0.1 s |
| AC magnet acceleration ramp | $4.4 \cdot 10^{-3}$ s |
| dB/dt (AC magnet) | 272 T/s |

Our future work includes as accurately as possible measurements of the test magnet cable cryogenic power losses. We also will upgrade the magnet power supply to double the discharge voltage of the capacitor bank. This will allow to increase operating current to 2 kA, and the B-field in magnet gaps up to 0.8 T. With such upgrade the dB/dt ramping rates of up to 600 T/s should be achievable allowing to determine cryogenic power losses with higher precision and make projection of required cryogenic support for the future large accelerators more reliable.

## ACKNOWLEDGMENT

We would like to thank Frank McConologue for thoughtful engineering designs and Thomas Lynn for meticulous magnet test system assembly work. We are also grateful to Jamie Blowers for the HTS magnet B-field simulations.